%% file: main.tex
\documentclass{article}

\usepackage{hyperref}
\hypersetup{
	colorlinks   = true, 
	urlcolor     = blue, 
	linkcolor    = blue, 
	citecolor   = blue 
}

\usepackage[caption=false,font=footnotesize]{subfig}

\usepackage[utf8]{inputenc}
\usepackage{amsmath}
\usepackage{amssymb}
\usepackage{ifthen}

\usepackage{graphicx}

\newcommand{\compilefullversion}{true}
\ifthenelse{\equal{\compilefullversion}{false}}{%
	\newcommand{\OnlyInFull}[1]{}
	\newcommand{\OnlyInShort}[1]{#1}
}{%
	\newcommand{\OnlyInFull}[1]{#1}%
	\newcommand{\OnlyInShort}[1]{}%
}%

\OnlyInFull{
\usepackage{fullpage}
\usepackage{amsthm}
}

\newcommand{\E}{\mathbb{E}}
\newcommand{\I}{\mathbb{I}}

\newcommand{\cE}{\mathcal{E}}

\newcommand{\cR}{\mathcal{R}}

\newcommand{\OPT}{{\it OPT}}
\newcommand{\LB}{{\it LB}}

\newcommand{\rtheta}{\tilde{\theta}}
\newcommand{\esigma}{\hat{\sigma}}

\title{An Issue in the Martingale Analysis of \\ the Influence Maximization Algorithm IMM}

\OnlyInFull{\author{Wei Chen \\ Microsoft Research, Beijing, China \\ weic@microsoft.com }}

\OnlyInShort{\titlerunning{An Issue in the Martingale Analysis of IMM}
\author{Wei Chen}
\authorrunning{Wei Chen}
\institute{Microsoft Research, Beijing, China \\
	\email{weic@microsoft.com}}
}

\date{}

\begin{document}

\maketitle

\begin{abstract}
	This paper explains a subtle issue in the martingale analysis of the IMM algorithm, a
	state-of-the-art influence maximization algorithm.
	Two workarounds are proposed to fix the issue, both requiring minor changes on the algorithm
	and incurring a slight penalty on the running time of the algorithm.

\end{abstract}

\section{Introduction}

Tang et al. design a scalable influence maximization algorithm IMM (Influence Maximization
	with Martingales) in \cite{tang15}, and apply martingale inequalities to the analysis.
In this paper, we describe a subtle issue in their martingale-based analysis.
The consequence is that the current proof showing that the IMM algorithm guarantees 
	$(1-1/e-\varepsilon)$ approximation with high probability is technically incorrect.
We provide a detailed explanation about the issue, and further propose two
	possible workarounds to address the issue, but both workarounds require
	minor changes to the algorithm with a slight penalty on running time.
Xiaokui Xiao, one of the authors of \cite{tang15}, has acknowledged the issue pointed out in this paper.
	
\subsection{Background and Related Work}
Influence maximization is the problem of given a social network $G=(V,E)$, 
	a stochastic diffusion model
	with parameters on the network, and a budget of $k$ seeds, finding the optimal
	$k$ seeds $S\subseteq V$ such that the influence spread of the seeds $S$, denoted
	as $\sigma(S)$ and 
	defined as the expected number of nodes activated based on diffusion model starting from
	$S$, is maximized.
The influence maximization is originally formulated as a discrete optimization problem
	by Kempe et al.~\cite{kempe03}, and has been extensively studied
	in the literature (cf. \cite{chen2013information} for a survey).
One important direction is scalable influence 
	maximization~\cite{ChenWY09,ChenWW10,ChenYZ10,simpath,JungHC12,BorgsBrautbarChayesLucier,CohenDPW14,tang14,tang15,NguyenTD16}, which focuses on improving the efficiency of running
	influence maximization algorithms on large-scale networks.
The early studies on this direction are heuristics based on graph 
	algorithms~\cite{ChenWY09,ChenWW10,ChenYZ10,simpath,JungHC12} or 
	sketch-based algorithms~\cite{CohenDPW14}.
Borgs et al. propose the novel reverse influence sampling (RIS) approach, which achieves 
	theoretical guarantees on both the approximation ratio and near-linear expected running 
	time~\cite{BorgsBrautbarChayesLucier}.
The RIS approach is further improved in~\cite{tang14,tang15,NguyenTD16} to achieve 
	scalable performance on networks with billions of nodes and edges.
The IMM algorithm we discuss in this paper is from~\cite{tang15}, which uses the martingales
	to improve the performance, and is considered as one of the state-of-the-art influence
	maximization algorithms.
However, we show in this paper that the algorithm has a subtle issue
	that affects its correctness.
The IMM algorithm has been used in later studies as a component (e.g.~\cite{YangMPH16,CT17,SHYC18}), so
	it is worth to point out the issue and the workarounds for the correct usage of the IMM
	algorithm.
The SSA/D-SSA algorithm of~\cite{NguyenTD16} is another state-of-the-art influence maximization
	algorithm, but the original publication also contains several analytical issues, which have been pointed out in~\cite{HuangWBXL17}.

\section{Description of the Issue}

\subsection{Brief Description of the RIS Approach}

At the core of the RIS approach is the concept of reverse-reachable (RR) sets.
Given a network $G=(V,E)$ and a diffusion model, an RR set $R \subseteq V$ is 
	sampled by first randomly selecting a node $v\in V$ and then 
	reverse simulating the diffusion process
	and adding all nodes reached by the reverse simulation into $R$.
Such reverse simulation can be carried out efficiently for a large class of diffusion
	models called the
	triggering model (see~\cite{kempe03,tang15} for model details).
Intuitively, each node $u\in R$ if acting as a seed would activate $v$ in the corresponding
	forward propagation, and based on this intuition
	the key relationship $\sigma(S) = n\cdot \E [\I\{S \cap R \ne \emptyset \}]$
	is established, where $\sigma(S)$ is the influence spread, $n = |V|$, and $\I$ is the indicator function.
The RIS approach is to collect enough number of RR sets $\cR = \{R_1, R_2, \ldots, R_\theta \}$,
	so that $\sigma(S)$ can be approximated by 
	$\esigma(S) = n \cdot \sum_{i=1}^\theta \I\{R_i \cap S \ne \emptyset \} /\theta $.
We call $R_i \cap S \ne \emptyset $ as $S$ covering $R_i$.
Thus, the original influence maximization problem is converted to finding $k$ seeds $S$ 
	that can cover the most number of RR sets in $\cR$.
This is a $k$-max coverage problem, and a greedy algorithm 
	(referred to as the NodeSelection procedure in IMM~\cite{tang15})
	can be applied to solve it
	with a $1-1/e$ approximation ratio.
	
\begin{figure}[t] 
	\centering
	\caption{Algorithm 2 (Sampling procedure) of IMM as in the original paper~\cite{tang15}.}
	{\includegraphics[width=0.7\linewidth]{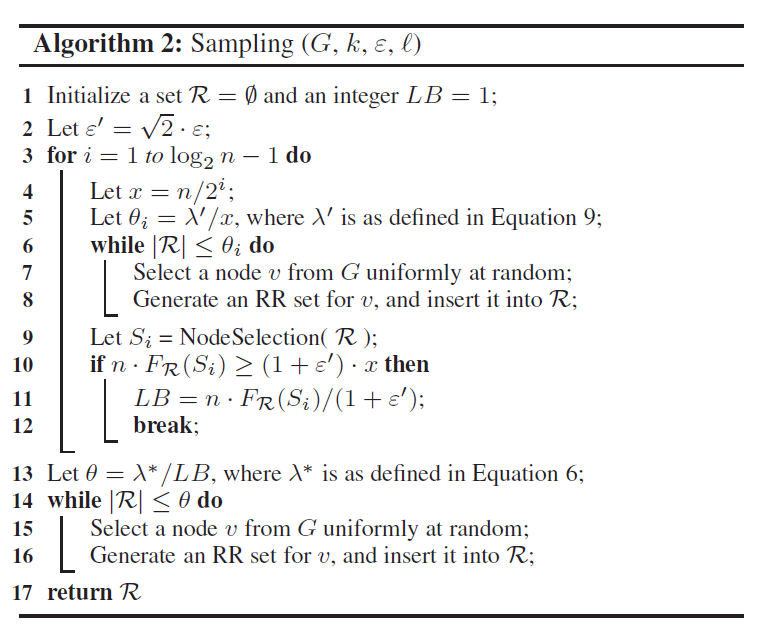}\label{fig:sampling}}
\end{figure}

Implementations of the RIS approach differ in their estimation of the number of RR sets needed.
IMM algorithm~\cite{tang15} iteratively doubles the number of RR sets until it obtains a reasonable estimate
	$\LB$ as the lower bound of the optimal solution $\OPT$, and then apply a formula
	$\theta = \lambda^*/\LB$, where $\lambda^*$ is a constant dependent on the problem instance,
	to get the final number of RR sets needed
	(See Fig.~\ref{fig:sampling} for the reprint of the Sampling procedure of IMM).

\subsection{Summary of the Issue}
\label{sec:issuesummary}

The main issue of the IMM analysis in~\cite{tang15} 
	is at its correctness claim of Theorem 4, which shows that 
	the output of IMM gives a $1-1/e-\varepsilon$ approximate solution with probability
	at least $1 - 1/n^\ell$.
The proof of this part is very brief, containing only one sentence as excerpted below, which
combines the result from Theorem 1 and Theorem 2.
\\

``{\em By combining Theorems 1 and 2, we obtain
	that Algorithm 3 returns a $(1-1/e - \varepsilon)$-approximate solution
	with at least $1-1/n^\ell$ (probability).}''	\\

At the high level, Theorem 1 claims that if NodeSelection procedure is fed with 
an RR set sequence of length at least $\theta \ge \lambda^* / \OPT$, then 
with probability at least $1 - 1/n^\ell$, NodeSelection outputs a seed set that
is a $1-1/e-\varepsilon$ approximate solution.
Then Theorem 2 claims that the Sampling procedure outputs an RR set sequence of length
at least $\lambda^* / \OPT$ with probability at least $1 - 1/n^\ell$.
It may appear that we could use a simple union bound to combine the two theorems
	to show that IMM achieves the $1-1/e-\varepsilon$ approximation with
	probability at least $1 - 2/n^\ell$.
%
%
Finally, we just need to reset $\ell = \ell + \log 2/ \log n$ to change the probability from
$1 - 2/n^\ell$ to $1 - 1/n^\ell$.\footnote{The original paper has a typo here.
	It says to reset $\ell$ to $\ell(1+\log 2/\log n)$, but this is not necessary.
	Only resetting $\ell$ to $\ell + \log 2/ \log n$ is enough.}

However, with a closer inspection, Theorem 1 is true only for each {\bf fixed length}
$\theta \ge \lambda^* / \OPT$, but the Sampling procedure returns an RR set sequence of 
	{\bf random length}.
Henceforth, to make the distinction explicit, we use $\rtheta$ to denote the random length
	returned by the Sampling procedure.
Technically, this $\rtheta$ is a {\em stopping time}, a concept frequently used in 
	martingale processes~\cite{MU05}.
Thus, what Theorem 2 actually claims is $\Pr\{\rtheta \ge \lambda^* / \OPT \} \ge 1 - 1/n^\ell $.
Due to this discrepancy between fixed length and random length in RR set sequences, 
	  we cannot directly combine Theorem 1 and Theorem 2 to obtain Theorem 4 as in the paper.
This is the main issue of the analysis in the IMM paper~\cite{tang15}.

%

In the next two subsections, we will provide more detailed discussion to illustrate the above
issue.
In Section~\ref{sec:probspace}, we first make it explicit what is the exact probability space 
we use for the analysis of the IMM algorithm.
Then in Section~\ref{sec:details}, we go through lemma by lemma on the original analysis
to make the distinction between the fixed length $\theta$ and the random stopping time
$\rtheta$ explicit, so that the issue summarized above is more clearly illustrated.

\subsection{Treatment on the Probability Space}
\label{sec:probspace}

For the following discussion, we will frequently refer to certain details in the
	Sampling procedure of IMM, namely Algorithm 2 of IMM in~\cite{tang15}
	(see Fig.~\ref{fig:sampling}).

To clearly understand the random stopping time $\rtheta$, we first clarify the probability space
	upon which $\rtheta$ is defined.
We first note that from the algorithm, 
	the maximum possible number of RR sets the algorithm could generate is
	$\lceil \lambda^* \rceil$ (defined in Eq.(6)).
Thus we view the probability space as the space of all $\lceil \lambda^* \rceil$ RR set sequences 
	$R_1, R_2, \ldots, R_{\lceil \lambda^* \rceil}$, where each $R_i$ is generated i.i.d. 
We denote this space as $\Omega$.
Then in one run of the IMM algorithm, one such RR set sequence $\cR_0$ 
	is drawn from the probability space $\Omega$. 
In the $i$-th iteration of the Sampling procedure, the algorithm gets the prefix of the 
	first $\theta_i$ ($\theta_i$ is defined in
	line 5 of Algorithm 2) RR sets in the above sequence $\cR_0$, and based on certain condition
	about this prefix 
	the algorithm decides whether to continue the iteration or stop; and when it stops,
	it determines the final number $\rtheta = \lambda^*/\LB$ of RR sets needed, and retrieves the prefix of
	$\rtheta$ RR sets from $\cR_0$.
Note that $\rtheta$ here is the $\theta$ used in line 13 of Algorithm 2, but we explicitly use
	$\rtheta$ to denote that it is a random variable (because
	$\LB$ is a random variable), and its value is determined by
	the prefix of RR sets in $R_1, R_2, \ldots$.
In contrast, for a fixed $\theta$ such as the $\theta$ used in Theorem 1, it simply corresponds
	to the $\theta$ RR sets in the sequence sample $\cR_0$.
For convenience, we use $\cR_0[\theta]$ to denote the prefix of $\cR_0$ of fixed length $\theta$,
and $\Omega[\theta]$ to be the subspace of all RR set sequences of length $\theta$.
Note that we use $\Omega$ and $\Omega[\theta]$ to refer to both the set of sequences and
their distribution.

\subsection{Detailed Discussion by Revisiting All Lemmas and Theorems}
\label{sec:details}

Hopefully we clarify the distinction between the fixed-length sequence $R_1, R_2$, $\ldots, R_\theta$ and
	the actual sequence $R_1, R_2, \ldots, R_{\rtheta}$ generated by the sampling phase with
	a random stopping time $\rtheta$.
We now revisit the technical lemmas and the theorems of the paper to explicitly
	distinguish between the usage of fixed length $\theta$ and random length $\rtheta$.

First and foremost, the martingale
	inequalities summarized in Corollaries 1 and 2 should only work for a fixed constant $\theta$, 
	not for a random stopping time, because they come from standard martingale
	inequalities as summarized in~\cite{CL06}, which deals with martingales of fixed length.
However, the authors introduce these inequalities in the context of RR set sequence generated
	by the Sampling procedure (see the first sentence in Section 3.1 of~\cite{tang15}).
As we explained, the RR set sequence generated by the Sampling procedure has random length
	$\rtheta$, so Corollaries 1 and 2 should not be applied to such random length sequences.
This is the source of confusion leading to the incorrectness of the proof of Theorem 4.
Henceforth, we should clearly remember that Corollaries 1 and 2 only work for fixed length
	$\theta$.

Next, for Lemmas 3 and 4, the $\theta$ there should refer to a fixed number,
	because their proofs rely on the martingale inequalities in Corollary 1 and 2, which
	are correct only for a fixed $\theta$.

For Theorem 1, same as discussed above, if we view $\theta$ as a fixed constant, then 
	Theorem 1 is correct.
We need to remark here that Theorem 1 talks about the node selection phase, so its exact
meaning is that if we feed the NodeSelection procedure with an RR set sequence
of {\bf fixed length} $\theta$, randomly drawn from the space $\Omega[\theta]$,
then the node selection phase would return an approximate
solution.
Therefore, it is not applicable when the NodeSelection procedure is fed with the RR set
	sequence generated from the Sampling procedure, since this sequence has a random length
	and is not drawn from the space $\Omega[\theta]$ for a fixed $\theta$.
	
%

Lemma 5 and Corollary 3 are still correct, since they are not related to the application of
	martingale inequality.
For Lemmas 6 and 7, again they are correct when $\theta$ is a fixed number satisfying inequality (8).

For Theorem 2, as already mentioned in Section~\ref{sec:issuesummary}, it is about
	the RR set sequence $\cR = \{R_1, R_2, \ldots, R_{\rtheta} \}$ generated by
	the Sampling procedure, with random length $\rtheta$, and its
	technical claim is 
\begin{equation} \label{eq:theorem2}
\Pr \left\{ \rtheta \ge \frac{\lambda^*}{\OPT} \right\} \ge 1 - \frac{1}{n^{\ell}},
\end{equation}
where the probability is taken
	from the probability space $\Omega$, the random sample
	$\cR_0$ of which determines the actual random length of output
	$\rtheta$.
The proof of Theorem 2 uses Lemma 6 and Lemma 7. 
When it uses Lemma 6 and Lemma 7, it is in the context of the Sampling procedure, 
	and the $\theta$ used for
	Lemma 6 and Lemma 7 in this context is exactly the $\theta_i = \lambda' / x_i$ defined in 
	line 5 of algorithm, where $\lambda'$ is a constant defined in Eq.(9), and $x_i = n/2^i$, and
	$i$ refers to the $i$-th iteration in the Sampling procedure.
Therefore, $\theta_i$ indeed is a constant that does not depend on the generated RR sets, and
	the applications of Lemmas 6 and 7 is in general appropriate.
However, the original proof of Theorem 2 is brief, and there is a subtle point that may not be
	clear from the proof, and thus some extra clarification is deserved here.

The subtlety is that, Lemmas 6 and 7 are correct when the NodeSelection procedure is fed
	with a fixed length RR set sequence sampled from $\Omega[\theta]$.
However, in the $i$-th iteration of the Sampling procedure, the actual RR set sequence
	fed into NodeSelection is {\em not} sampled from the space $\Omega[\theta_i]$.
This is because
	the fact that the algorithm enters the $i$-th iteration implies that 
	the previous RR set sequence failed the coverage condition check in line 10 in the
	previous iterations, and thus the actual sequence fed into NodeSelection in the $i$-th
	iteration is a biased sample.
This subtlety makes the rigorous proof of Theorem 2 longer, but does not invalidate the
	Theorem.
Intuitively, for a random sample $\cR_0[\theta_i]$ drawn from $\Omega[\theta_i]$, even if
	$\cR_0[\theta_i]$ would not make the algorithm survive to the $i$-th iteration, 
	we could still treat it as if it is fed to NodeSelection in the $i$-th iteration, and
	use Lemmas 6 and 7 to argue that some event $\cE_i$ only occurs with a small probability
	$\delta_3$.
Then the event that both algorithm enters the $i$-th iteration and  $\cE_i$ occurs
	must be also smaller than $\delta_3$.
\OnlyInFull{For completeness, 
		in the appendix, we provide a more rigorous technical proof of Theorem 2 applying the above idea.}\OnlyInShort{For completeness, 
		in~\cite{Chen18}, we provide a more rigorous technical proof of Theorem 2 applying the above idea.}

Continuing to Lemmas 8 and 9, similar to Lemma 6 and Lemma 7, it is correct when we treat $\theta$ 
	as a constant.
For Lemma 9, it uses Lemma 8, and if we treat the application of Lemma 8 in the same way
	as we treat the application of Lemmas 6 and 7 in the proof of Theorem 2, then 
	Lemma 9 is correct.
Lemma 10 and Theorem 3 are independent of the application of martingale inequalities
	and are correct.

Finally, we investigate the proof of Theorem 4, in particular the part on the correctness of
	the IMM algorithm.
As outlined in Section~\ref{sec:issuesummary}, a direct combination of Theorem 1 and Theorem 2
	is problematic.
We now discuss this point with more technical details.
	
%
%
%
%
For Theorem 1, based on our above discussion, it works for a fixed value of $\theta$. 
More precisely, when we use the setting discussed after Theorem 1, what it really says is that,
for all fixed $\theta \ge \lambda^*/\OPT$, if we use a random sample $\cR_0[\theta]$ drawn from
	distribution $\Omega[\theta]$, then when we feed the NodeSelection procedure with
	$\cR_0[\theta]$, the probability that NodeSelection returns a seed set that is a
	$(1-1/e-\varepsilon)$ approximate solution is at least $1 - 1/n^\ell$.
To make it more explicit, let $S^*_k(\cR)$ be the seed set returned by NodeSelection under
	input RR set sequence $\cR$.
Let $Y(S)$ be an indicator, and it is $1$ when seed set $S$ is a $(1-1/e-\varepsilon)$ approximate solution, and it is $0$ otherwise.
Then, what Theorem 1 says is,
\begin{equation} \label{eq:theorem1}
\forall \theta \ge \lambda^*/\OPT,
\Pr_{\cR_0[\theta]\sim \Omega[\theta]}\{ Y(S^*_k(\cR_0[\theta])) = 1 \} \ge 1 - \frac{1}{n^\ell}.
\end{equation}
	
Next, as discussed above, what Theorem 2 really says is given in Eq.~\eqref{eq:theorem2}.
Also to make it more precise and use the same base sample from the probability space, let
	$\cR_0$ be the sample drawn from $\Omega$, and let
	$\cR(\cR_0) = \{R_1, R_2, \ldots, R_{\rtheta} \}$ be the sequence generated by the Sampling 
	procedure, and $\rtheta(\cR_0)$ denote its length.
Thus by definition, $\cR(\cR_0)$ is the first $\rtheta(\cR_0)$ RR sets of $\cR_0$.
Then Theorem 2 (and Eq.~\eqref{eq:theorem2}) is restated as
\begin{equation} \label{eq:theorem22}
\Pr_{\cR_0 \sim \Omega} \left\{ \rtheta(\cR_0) \ge \frac{\lambda^*}{\OPT} \right\} \ge 1 -\frac{1}{n^{\ell}}.
\end{equation}
For Theorem 4, we want to bound the probability that using the Sampling procedure output
	$\cR(\cR_0)$ to feed into NodeSelection, its output fails to provide the 
	$1-1/e-\varepsilon$ approximation ratio, that is, 
\begin{equation}
\label{eq:thm4}
\Pr_{\cR_0\sim \Omega}\{ Y(S^*_k(\cR(\cR_0))) = 0 \} \le \frac{2}{n^\ell}.
\end{equation}

The following derivation further separates the left-hand side of Eq.~\eqref{eq:thm4} into
	two parts by the union bound:

\begin{align}
&\Pr_{\cR_0\sim \Omega}\{ Y(S^*_k(\cR(\cR_0))) = 0 \} \nonumber \\
& \le \Pr_{\cR_0\sim \Omega} \left\{ \rtheta(\cR_0) < \frac{\lambda^*}{\OPT} \vee 
		\left (\rtheta(\cR_0) \ge \frac{\lambda^*}{\OPT} \wedge Y(S^*_k(\cR(\cR_0))) = 0 \right) \right\} \nonumber \\
& \le \Pr_{\cR_0\sim \Omega} \left\{ \rtheta(\cR_0) < \frac{\lambda^*}{\OPT} \right\}
	+ \Pr_{\cR_0\sim \Omega} \left\{ \rtheta(\cR_0) \ge \frac{\lambda^*}{\OPT} \wedge Y(S^*_k(\cR(\cR_0))) = 0 \right\} \nonumber \\
& \le \frac{1}{n^{\ell}} + \Pr_{\cR_0\sim \Omega} \left\{ \rtheta(\cR_0) \ge \frac{\lambda^*}{\OPT} \wedge Y(S^*_k(\cR(\cR_0))) = 0 \right\}, \label{eq:separateThm4}
\end{align}
where the last inequality is by Theorem 2 (Eq.~\eqref{eq:theorem22}).
To continue, we want to bound 
\begin{align}
\Pr_{\cR_0\sim \Omega} \left\{ \rtheta(\cR_0) \ge \frac{\lambda^*}{\OPT} \wedge Y(S^*_k(\cR(\cR_0))) = 0 \right\} \le \frac{1}{n^\ell}. \label{eq:wantboundNodeSelect}
\end{align}
However, the above inequality is incompatible with Inequality~\eqref{eq:theorem1}, because Inequality~\eqref{eq:theorem1}
holds for {\bf each fixed} $\theta \ge \frac{\lambda^*}{\OPT}$, but 
Inequality~\eqref{eq:wantboundNodeSelect} is for {\bf all} 
$\rtheta(\cR_0) \ge \frac{\lambda^*}{\OPT}$.
This is where the direct combination of Theorem 1 and Theorem 2 would fail to produce
	the correctness part of Theorem 4.

\section{Possible Workarounds for the Issue}

It is unclear if the analysis could be fixed without changing any aspect of the algorithm. 
In this section, we propose two possible workarounds, both of which require at least some change
	to the algorithm and incur some running time penalty.
	
\subsection{Workaround 1: Regenerating New RR Sets}

One simple workaround is that in the IMM algorithm, after determining the final length 
	$\rtheta$ of the RR set sequence, regenerate the entire RR set sequence of length $\rtheta$
	from scratch, and use the newly generated sequence as the output of the Sampling algorithm
	and feed it into the final call to NodeSelection.
That is, after line 13 of Algorithm 2, regenerate $\rtheta$ RR sets instead of lines 14-16.

Intuitively, this would feed the final call of NodeSelection with an unbiased RR set sequence
	so that Theorem 1 can be applied.
We represent this new unbiased sequence as a new independent sample $\cR'_0$ from the probability
	space $\Omega$, and then taking the prefix of $\cR'_0$ with 
	$\rtheta(\cR_0)$ RR sets, where $\rtheta(\cR_0)$ is the number of RR sets determined from
	sequence $\cR_0$ that is needed for the final call of NodeSelection.
Thus we use the notation $\cR'_0[\rtheta(\cR_0)]$ to represent the RR set sequence that
	is fed into the final call of NodeSelection.
The correctness can be rigorously proved as follows.
First, Eq.~\eqref{eq:thm4} for Theorem 4 is changed to:
\begin{equation}
\label{eq:newthm4}
\Pr_{\cR_0\sim \Omega, \cR'_0\sim \Omega}\{ Y(S^*_k(\cR'_0[\rtheta(\cR_0)])) = 0 \} \le \frac{2}{n^\ell}.
\end{equation}
To show the above inequality, following a similar derivation as in Eq.~\eqref{eq:separateThm4}, what we need
	to show is the following instead of Eq.~\eqref{eq:wantboundNodeSelect}:
\begin{align}
\Pr_{\cR_0\sim \Omega, \cR'_0\sim \Omega} \left\{ \rtheta(\cR_0) \ge \frac{\lambda^*}{\OPT} \wedge Y(S^*_k(\cR'_0[\rtheta(\cR_0)])) = 0 \right\} \le \frac{1}{n^\ell}. \label{eq:wantboundNodeSelectnew}
\end{align}
This can be achieved by the following derivation:
\begin{align}
&\Pr_{\cR_0\sim \Omega, \cR'_0\sim \Omega} \left\{ \rtheta(\cR_0) \ge \frac{\lambda^*}{\OPT} \wedge Y(S^*_k(\cR'_0[\rtheta(\cR_0)])) = 0 \right\}  \nonumber \\
&= \Pr_{\cR_0\sim \Omega, \cR'_0\sim \Omega} 
\left\{ \bigvee_{\theta=\lceil \frac{\lambda^*}{\OPT} \rceil}^{\lceil \lambda^* \rceil }
\rtheta(\cR_0)  = \theta \wedge Y(S^*_k(\cR'_0[\rtheta(\cR_0)])) = 0
\right\} \nonumber \\
& \le \sum_{\theta=\lceil \frac{\lambda^*}{\OPT} \rceil}^{\lceil \lambda^* \rceil }
\Pr_{\cR_0\sim \Omega, \cR'_0\sim \Omega} 	\left\{ 
\rtheta(\cR_0)  = \theta \wedge Y(S^*_k(\cR'_0[\rtheta(\cR_0)])) = 0
\right\} \quad \quad  \mbox{\{union bound\}}
\nonumber \\
& =  \sum_{\theta=\lceil \frac{\lambda^*}{\OPT} \rceil}^{\lceil \lambda^* \rceil }
\Pr_{\cR_0\sim \Omega, \cR'_0\sim \Omega} 	\left\{ 
\rtheta(\cR_0)  = \theta \wedge Y(S^*_k(\cR'_0[\theta])) = 0
\right\}  \nonumber \\
& = \sum_{\theta=\lceil \frac{\lambda^*}{\OPT} \rceil}^{\lceil \lambda^* \rceil }
\Pr_{\cR_0\sim \Omega}\{\rtheta(\cR_0)  = \theta \} \cdot \Pr_{\cR'_0\sim \Omega} 	\left\{ Y(S^*_k(\cR'_0[\theta])) = 0 \right\}  
\quad \mbox{\{independence of $\cR_0$ and $\cR'_0$\}}
\label{eq:independentRRset}  \\
& = \sum_{\theta=\lceil \frac{\lambda^*}{\OPT} \rceil}^{\lceil \lambda^* \rceil }
\Pr_{\cR_0\sim \Omega}\{\rtheta(\cR_0)  = \theta \} \cdot \Pr_{\cR'_0[\theta]\sim \Omega[\theta]} 	\left\{ Y(S^*_k(\cR'_0[\theta])) = 0 \right\}  \nonumber \\
& \le \sum_{\theta=\lceil \frac{\lambda^*}{\OPT} \rceil}^{\lceil \lambda^* \rceil }
\Pr_{\cR_0\sim \Omega}\{\rtheta(\cR_0)  = \theta \} \cdot  \frac{1}{n^\ell}  
\quad  \quad  \quad \quad \quad  \quad  \quad  \quad \quad  \quad \mbox{\{Eq.~\eqref{eq:theorem1} of Theorem 1\}}
\nonumber \\
& = \frac{1}{n^\ell}. 
\nonumber
\end{align}
The key step is Eq.~\eqref{eq:independentRRset}, where because $\cR'_0$ is independent of
	$\cR_0$ (we regenerate a new RR set sequence for the last call to NodeSelection), we can represent the
	probability $\Pr_{\cR_0\sim \Omega, \cR'_0\sim \Omega} 	\left\{ 
	\rtheta(\cR_0)  = \theta \wedge Y(S^*_k(\cR'_0[\theta])) = 0
	\right\} $ as the product of two separate factors.
Therefore, the correctness part of Theorem 4 now holds. 
Note that within the Sampling procedure, we do not need to regenerate RR set sequences from 
	scratch (before line 9 of Algorithm 2), because by our detailed discussion in
	Section~\ref{sec:details}, even without regenerating RR sets, Theorem 2 still holds
	with a more careful argument.
	
In terms of the running time, this workaround at most doubles the number of RR sets generated,
	and thus its running time only adds a multiplicative factor of $2$ to the original
	result.
Therefore, the asymptotic running time remains
	as $O((k+\ell)(n+m)\log n / \varepsilon^2)$ in expectation.

\subsection{Workaround 2: Apply Union Bounding with Larger $\ell$}

The second workaround is by directly bounding Eq.~\eqref{eq:wantboundNodeSelect} by
	 a union bound, as shown in the derivation below.
\begin{align}
&\Pr_{\cR_0\sim \Omega} \left\{ \rtheta(\cR_0) \ge \frac{\lambda^*}{\OPT} \wedge Y(S^*_k(\cR(\cR_0))) = 0 \right\}  \nonumber \\
&= \Pr_{\cR_0\sim \Omega} 
\left\{ \bigvee_{\theta=\lceil \frac{\lambda^*}{\OPT} \rceil}^{\lceil \lambda^* \rceil }
\rtheta(\cR_0)  = \theta \wedge Y(S^*_k(\cR(\cR_0))) = 0
\right\} \nonumber \\
& \le \sum_{\theta=\lceil \frac{\lambda^*}{\OPT} \rceil}^{\lceil \lambda^* \rceil }
\Pr_{\cR_0\sim \Omega} 	\left\{ 
\rtheta(\cR_0)  = \theta \wedge Y(S^*_k(\cR(\cR_0))) = 0
\right\} \nonumber \\
& =  \sum_{\theta=\lceil \frac{\lambda^*}{\OPT} \rceil}^{\lceil \lambda^* \rceil }
\Pr_{\cR_0\sim \Omega} 	\left\{ 
\rtheta(\cR_0)  = \theta \wedge Y(S^*_k(\cR_0[\theta])) = 0
\right\}  \nonumber \\
& \le \sum_{\theta=\lceil \frac{\lambda^*}{\OPT} \rceil}^{\lceil \lambda^* \rceil }
\Pr_{\cR_0\sim \Omega} 	\left\{ 
Y(S^*_k(\cR_0[\theta])) = 0
\right\}  \label{eq:ignoretheta} \\
& = \sum_{\theta=\lceil \frac{\lambda^*}{\OPT} \rceil}^{\lceil \lambda^* \rceil }
\Pr_{\cR_0[\theta]\sim \Omega[\theta]} 	\left\{ 
Y(S^*_k(\cR_0[\theta])) = 0
\right\}  \nonumber \\
& \le \sum_{\theta=\lceil \frac{\lambda^*}{\OPT} \rceil}^{\lceil \lambda^* \rceil }
\frac{1}{n^\ell} & \mbox{\{by Theorem 1, Eq.~\eqref{eq:theorem1}\}} \\
& \le \frac{\lceil \lambda^* \rceil}{n^\ell}. \label{eq:newresult}
\end{align}
Comparing the above derivation with the similar one for workaround 1, the key difference
	is between Eq.~\eqref{eq:independentRRset}  and Eq.~\eqref{eq:ignoretheta}.
In Eq.~\eqref{eq:independentRRset}, we could keep 
	$\Pr_{\cR_0\sim \Omega}\{\rtheta(\cR_0)  = \theta \}$ because the event
	$\{\rtheta(\cR_0)  = \theta \}$ is independent of the event
	$\{Y(S^*_k(\cR'_0[\theta])) = 0 \} $ in the second term.
But in Eq.~\eqref{eq:ignoretheta}, we cannot extract 
	$\Pr_{\cR_0\sim \Omega}\{\rtheta(\cR_0)  = \theta \}$ because the event
	$\{\rtheta(\cR_0)  = \theta \}$ is correlated with the event
	$\{Y(S^*_k(\cR_0[\theta])) = 0 \} $ in the second term.
Thus we have to simply drop the event $\{\rtheta(\cR_0)  = \theta \}$, causing the bound
	to be inflated by a factor of $\lceil \lambda^* \rceil$.

Using Inequality~\eqref{eq:newresult}, our second workaround 
	is to enlarge $\ell$ to $\ell'$ so that $ \lceil \lambda^* \rceil/ n^{\ell'} \le 1/n^\ell$.
However, $\lambda^*$ is also dependent on $\ell$.
To make it clear, we write it as $\lambda^*(\ell)$.
What we want is to set $\ell' = \ell + \gamma$, such that
\begin{equation}
\frac{\lceil \lambda^*(\ell') \rceil}{n^{\ell'}} = 
\frac{\lceil \lambda^*(\ell + \gamma) \rceil}{n^{\ell + \gamma}} \le \frac{1}{n^\ell}.
\end{equation}
This means we want $\lceil \lambda^*(\ell + \gamma) \rceil \le n^\gamma$.
From Eqs.(5) and (6) in~\cite{tang15}, we have
\begin{align*}
\lambda^*(\ell) 
& = 2n \cdot \left((1-1/e)\cdot \sqrt{\ell\log n + \log 2} 
	+ \sqrt{(1-1/e)\cdot \left(\log \binom{n}{k} + \ell \log n + \log 2 \right)} \right)^2 \cdot \varepsilon^{-2} \\
& \le 8 n (k+\ell + 1) \log n \cdot \varepsilon^{-2} - 1,
\end{align*}
where the relaxation in the inequality above is loose, involving relaxing the first
	square root term to the second one, relaxing $(1-1/e)$ to $1$, relaxing $\binom{n}{k}$ to
	$n^k$, relaxing $\log 2$ to $\log n$, and thus the $-1$ above can be certainly compensated
	by the relaxation, and it is used for relaxing the $\lceil \lambda^*(\ell + \gamma) \rceil $
	next.
Thus, to achieve $\lceil \lambda^*(\ell + \gamma) \rceil \le n^\gamma$, 
	we just need $8 n (k+\ell +\gamma + 1) \log n \cdot \varepsilon^{-2} \le n^\gamma$.
Asymptotically, $\gamma > 1$ would be fine for large enough $n$.
For a conservative bound, it is very reasonable to assume that $\varepsilon^{-1} \le n$,
	$k+\ell+\gamma+1 \le n$, then we just need $8 \log n \le n^{\gamma - 4}$, which means
	setting $\gamma \ge 4 + \log (8\log n)/ \log n$ is enough.
Thus $\gamma$ is essentially a small constant.

In practice, $\gamma$ could be computed by a binary search 
	once the parameters $n$, $k$, $\ell$ and $\varepsilon$
	of the problem instance are given.
Then we can set $\ell = \ell + \log 2 / \log n + \gamma$ in the algorithm.
By increasing $\ell$ with a small constant $\gamma$ (e.g. $\gamma = 2.5$), the running
	time increases from $O(k + \ell)(m+n)\log n/\varepsilon^2)$ to
	 $O(k + \ell + \gamma)(m+n)\log n/\varepsilon^2)$, so the running time penalty is likely
	 to be smaller than that of the first workaround.
Our experimental results below validate this point.
	 
\subsection{Experimental Evaluation}
We evaluate the two workarounds and compare them against the original IMM algorithm on
	two real world datasets: (a) NetHEPT, a coauthorship network with 15233 nodes and 31373 edges,
	mined from arxiv.org high energy physics section, and (b) DBLP, another coauthorship network
	with 655K nodes and 1990K edges, mined from dblp.uni-trier.de.
We use independent cascade model with edge probabilities set by the weighted cascade
	method~\cite{kempe03}: edge $(u,v)$'s probability is $1/d_v$ where $d_v$ is the in-degree
	of $v$.
These datasets are frequently used in other influence maximization studies
	such as~\cite{ChenWW10,tang15,CT17}.
	
We use IMM, IMM-W1, and IMM-W2 to denote the original IMM, the IMM with the first and the second
	workarounds, respectively.
For IMM-W2, we use binary search to find an estimate of $\gamma$ satisfying
	$\lceil \lambda^*(\ell + \gamma) \rceil \le n^\gamma$.
We set parameters $\varepsilon = 0.1$, $\ell = 1$, and influence spread is the average of 
	$10000$ simulation runs. 
We test the algorithms in seed set sizes $k= 50, 100, \ldots, 500$.
The code is written in C++ and compiled by Visual Studio 2013, and is run on 
	a Surface Pro 4 with dual core 2.20GHz CPU and 16GB memory.

\begin{figure}[t]
	\centering
	\captionsetup[subfigure]{font=scriptsize,oneside,margin={0.6cm,0.0cm}}
	\setcounter{subfigure}{0}%
	\begin{tabular}{cc}
	\subfloat[influence spread, NetHEPT]
	{\includegraphics[width=0.4\linewidth]{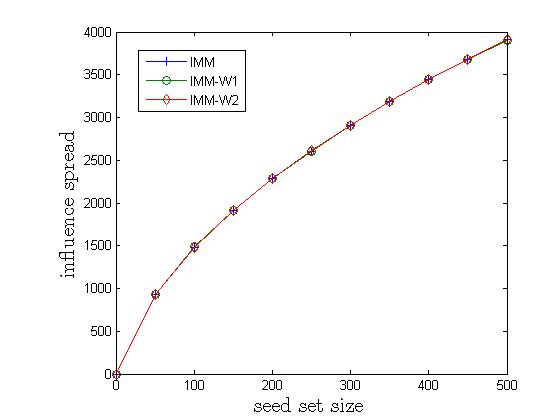}\label{fig:heptinf}} &
	\subfloat[influence spread, DBLP]
	{\includegraphics[width=0.4\linewidth]{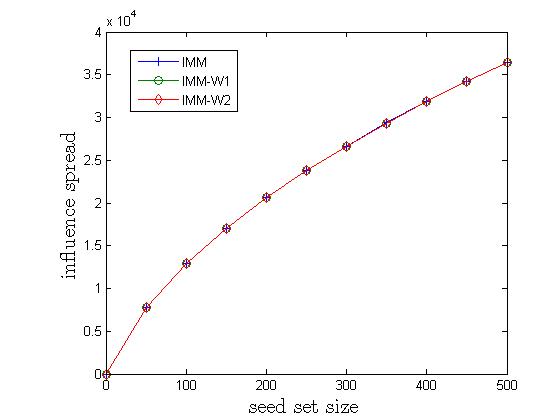}\label{fig:dblpinf}} \\
	\subfloat[running time, NetHEPT]
	{\includegraphics[width=0.4\linewidth]{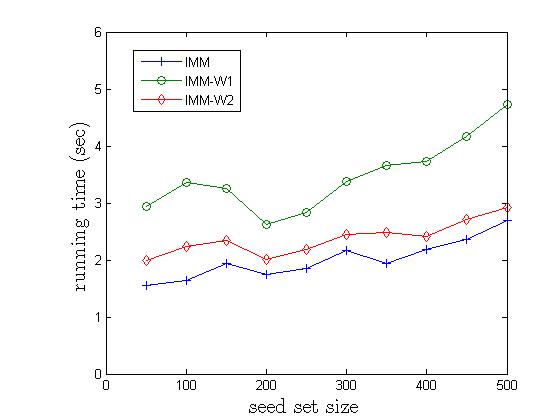}\label{fig:hepttime}} &
	\subfloat[running time, DBLP]
	{\includegraphics[width=0.4\linewidth]{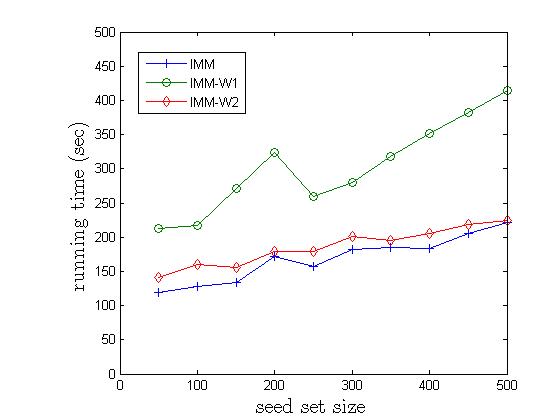}\label{fig:dblptime}}
	\end{tabular}
	\caption{Influence spread and running time results.}
	\label{fig:inftime}
\end{figure}


The influence spread and running time results are shown in Figure~\ref{fig:inftime}. 
As expected, all three algorithms achieve indistinguishable influence spread, since the
	two workarounds are to fix the theoretical issue on the dependency of RR sets, and should not
	affect much on the actual performance of the IMM algorithm.
In terms of running time, also as expected, IMM-W1 has the worst running time, but is within
	twice of running time of the IMM algorithm. 
IMM-W2 has much closer running time to IMM, though is still in general slower. 
We further observe that the $\gamma$ value used for IMM-W2 is within $2.5$ for the NetHEPT dataset
	and within $2$ for the DBLP dataset.
Therefore, it looks like that we can use the second workaround to provide a rigorous theoretical
	guarantee while achieving similar running time as the original IMM.

\section{Conclusion}
In this paper, we explain the issue in the original analysis of the IMM algorithm~\cite{tang15}.
Two workarounds are proposed, both of which require some minor changes to the algorithm and
	both incur a slight penalty in running time.
Since the IMM algorithm as a state-of-the-art influence maximization algorithm provides both
	strong theoretical guarantee and good practical performance, many follow-up studies
	in influence maximization use IMM algorithms as a template.
Thus, it is worth to point out this issue so that subsequent follow-ups will correctly use
	the algorithm, especially if they want to provide theoretical guarantee.
It remains an open question if the issue can be fixed without changing the original algorithm,
	or if a workaround with an even less impact to the algorithm and its running time can
	be found.

\section*{Acknowledgment}
The author would like to thank Jian Li for helpful discussions and verification on the issue
	explained in the paper.

\bibliography{bibdatabase} 
\bibliographystyle{plain}

\OnlyInFull{\input{appendix.tex}}

\end{document}

%% file: appendix.tex
\clearpage

\appendix

\section*{Appendix}

\section{Detailed Proof of Theorem 2 in \cite{tang15}}

In this appendix, we provide a detailed re-proof of Theorem 2 in \cite{tang15}. 
The proof follows the idea outlined in Section~\ref{sec:details}, and is meant to make the
	original proof of Theorem 2 in \cite{tang15} more rigorous and eliminate possible ambiguities
	on handling the martingale sequence dependency in the original proof.

\begin{proof}[Detailed Proof of Theorem 2]
	Let $x_i = n / 2^i$ for $i = 1, 2, \ldots, \lfloor \log_2 n \rfloor$, and
	$x_{\lfloor \log_2 n \rfloor+1} = 1$.
	Let $i$ be the index such that $ x_{i+1} \le \OPT < x_i$.
	Let $\theta_i = \lambda' / x_i$, where $\lambda'$ is defined in Eq.(9) of~\cite{tang15}.
	For a generic RR set sequence $\cR$, we use $S^*_k(\cR)$ to denote the output of the NodeSelection
	procedure with input $\cR$.
	Let $\cR_0$ denote the full same RR set sequence from the probability space $\Omega$
	that corresponds to a run of the IMM algorithm.
	Note that the $\cR$ appearing in lines 9 and 10 is the prefix of $\cR_0$ with length $\theta_i$, 
	so we use expression $\cR_0[\theta_i]$ explicitly to replace $\cR$ in these lines.
	This means the output $S_i$ in line 9 should be written as $S^*_k(\cR_0[\theta_i])$.
	Define event $\cE_i = \{ n \cdot F_{\cR_0[\theta_i]}(S^*_k(\cR_0[\theta_i])) \ge (1+\varepsilon')\cdot x_i \}$,
	which corresponds to the condition in line 10.
	Note that only when $\cE_1, \ldots, \cE_{i-1}$ are false, the Sampling procedure
		would enter the $i$-th iteration, and would check the condition in line 10
		corresponding to the even $\cE_i$.
	Let $\LB$ be the lower bound computed in line 11, just before the Algorithm 2 breaks the for-loop.
	
	To prove Theorem 2 (or equivalent Eq.~\eqref{eq:theorem2}), 
	by line 13 of Algorithm 2, it is sufficient to show that with low probability 
	$\LB > \OPT$.
	The way to show this is by showing that for each round $j \le i$, only with a low probability
	the algorithm breaks out of the for-loop in the $j$-th iteration; and
	if the algorithm breaks out of the for-loop at least in the $(i+1)$-th iteration or above,
	with a low probability, $\LB > \OPT$.
	Formally, we have
	\begin{align} 
	& \Pr \left\{ \rtheta \le \frac{\lambda^*}{\OPT} \right\} \nonumber \\
	& \le \Pr \left\{ \LB > \OPT \right\} \nonumber \\
	& \le \Pr \left\{ \bigvee_{j=1}^i \left( \left(\bigwedge_{s=1}^{j-1} \neg\cE_s \right) 
	\wedge \cE_j \right) \vee 
	\left( \left( \bigwedge_{j=1}^i \neg\cE_j \right) \wedge 
	(\LB > \OPT)  \right)\right\} \nonumber \\
	& \le \sum_{j=1}^i \Pr \left\{  \left(\bigwedge_{s=1}^{j-1} \neg\cE_s \right) 
	\wedge \cE_j \right\} + \Pr \left\{ \left( \bigwedge_{j=1}^i \neg\cE_j \right) \wedge 
	(\LB > \OPT)  \right\}. \label{eq:goodtheta}
	\end{align}
	
	For each $j\le i$, we want to bound $\Pr \left\{  \left(\bigwedge_{s=1}^{j-1} \neg\cE_s \right) 
	\wedge \cE_j \right\}$ by Lemma 6.
	As discussed above, Lemma 6 is for a fixed $\theta$ and a sample sequence $\cR$ drawn
	from $\Omega[\theta]$, but event $ \left(\bigwedge_{s=1}^{j-1} \neg\cE_s \right) 
	\wedge \cE_j $ means that the sequence $\cR[\theta_j]$ has passed through 
	iterations $1, 2, \ldots, j-1$.
	Thus, to connect with Lemma 6, we drop the events $\bigwedge_{s=1}^{j-1} \neg\cE_s$.
	Formally, we have
	\begin{align}
	& \Pr \left\{  \left(\bigwedge_{s=1}^{j-1} \neg\cE_s \right) 
	\wedge \cE_j \right\} \nonumber \\
	& \le \Pr \{ \cE_j \} \nonumber \\
	& = \Pr \{  n \cdot F_{\cR_0[\theta_j]}(S^*_k(\cR_0[\theta_j])) \ge (1+\varepsilon')\cdot x_j   \} \nonumber \\
	& \le \frac{1}{n^\ell \cdot \log_2 n}, \label{eq:earlybreak}
	\end{align}
	where the last inequality is by applying Lemma 6 with  $x = x_j$ and 
	$\delta_3 = 1/(n^\ell \cdot \log_2 n)$, since we
	have $j \le i$, which implies that $x_j \ge x_i > \OPT$.
	
	Next, we want to bound $\Pr \left\{ \left( \bigwedge_{j=1}^i \neg\cE_j \right) \wedge 
	(\LB > \OPT)  \right\}$ by applying Lemma 7.
	Note that event $\bigwedge_{j=1}^i \neg\cE_j$ implies that 
	the for-loop ends at some iteration $i^* > i$, and thus $\OPT \ge x_{i^*}$.
	If the algorithm never breaks the for-loop from line 12, that means $\LB=1$ and
	we have $\Pr \left\{ \left( \bigwedge_{j=1}^i \neg\cE_j \right) \wedge 
	(\LB > \OPT)  \right\} = 0$.
	So suppose the algorithm indeeds break from line 12 at some iteration $i^*$, 
	and by line 11 we have $\LB = n \cdot F_{\cR_0[\theta_{i^*}]}(S^*_k(\cR_0[\theta_{i^*}])) / (1+\varepsilon')$.
	Then we have
	\begin{align}
	& \Pr \left\{ \left( \bigwedge_{j=1}^i \neg\cE_j \right) \wedge (\LB > \OPT)  \right\} \nonumber \\
	& \le \Pr \left\{ (i^* \ge i+1) \wedge 
	(n \cdot F_{\cR_0[\theta_{i^*}]}(S^*_k(\cR_0[\theta_{i^*}])) / (1+\varepsilon') >\OPT)  \right\} \nonumber \\
	& = \Pr \left\{ \bigvee_{j=i+1}^{\lfloor \log_2 n \rfloor} (i^* =j ) \wedge 
	(n \cdot F_{\cR_0[\theta_{i^*}]}(S^*_k(\cR_0[\theta_{i^*}])) / (1+\varepsilon') >\OPT)  \right\} \nonumber \\
	& \le \sum_{j=i+1}^{\lfloor \log_2 n \rfloor} \Pr \left\{ (i^* =j ) \wedge 
	(n \cdot F_{\cR_0[\theta_{i^*}]}(S^*_k(\cR_0[\theta_{i^*}])) / (1+\varepsilon') >\OPT)  \right\} 
	& \mbox{\{union bound\}} \nonumber \\
	& \le \sum_{j=i+1}^{\lfloor \log_2 n \rfloor} \Pr \left\{ (i^* =j ) \wedge 
	(n \cdot F_{\cR_0[\theta_{j}]}(S^*_k(\cR_0[\theta_{j}])) / (1+\varepsilon') >\OPT)  \right\} \nonumber \\
	& \le \sum_{j=i+1}^{\lfloor \log_2 n \rfloor} \Pr \left\{ 
	(n \cdot F_{\cR_0[\theta_{j}]}(S^*_k(\cR_0[\theta_{j}])) / (1+\varepsilon') >\OPT)  \right\} \nonumber \\
	& \le \frac{\lfloor \log_2 n \rfloor - i}{n^\ell \cdot \log_2 n},
	& \mbox{\{by Lemma 7\}} \label{eq:LB}
	\end{align}
	where the last inequality is by Lemma 7 with $x = x_{j}$ and $\delta_3 = 1/(n^\ell \cdot \log_2 n)$,
	since $\OPT \ge x_{i+1} \ge x_j$.
	Finally, Theorem 2 is proved by combining Eqs.~\eqref{eq:goodtheta}, \eqref{eq:earlybreak},
	and~\eqref{eq:LB}:
	if $ i < \lfloor \log_2 n \rfloor$, then applying the above three inequalities;
	if $i = \lfloor \log_2 n \rfloor$, then we know that $\Pr \left\{ \left( \bigwedge_{j=1}^i \neg\cE_j \right) \wedge 
	(\LB > \OPT)  \right\} = 0$, so applying Eqs.~\eqref{eq:goodtheta} and \eqref{eq:earlybreak}
	are enough.
\end{proof}